\documentclass[pra,twocolumn,superscriptaddress]{revtex4-2} 
\usepackage{mathrsfs}
\usepackage{amsfonts}
\usepackage{amsmath}
\usepackage{txfonts}
\usepackage{amssymb}
\usepackage{graphicx,subfigure}
\usepackage{bm}
\usepackage{color}
\usepackage[normalem]{ulem}
\usepackage{dcolumn} 
\usepackage[table]{xcolor}
\usepackage{float}
\usepackage{tabularx}
\usepackage{hyperref}
\usepackage{etoolbox}
\usepackage{etoolbox}
\newcommand{\ket}[1]{|#1\rangle}
\newcommand{\bra}[1]{\langle #1|}

\begin{document}
\title{Geometric phases of reduced states in the transverse-field Ising chain} 
\author{Chiragkumar R. Vasani}
\email{chiragvasani090@gmail.com}
\author{Erik Sj\"oqvist}
\email{erik.sjoqvist@physics.uu.se}
\affiliation{Department of Physics and Astronomy, Uppsala University, 
Box 524, SE-751 20 Uppsala, Sweden}
\date{\today}
\begin{abstract}
Geometric phases have been extensively investigated in a wide range of quantum systems, often revealing deep connections to the underlying topology of many-body states. In this work, we examine two geometric phases defined for mixed quantum states—the interferometric geometric phase and the Uhlmann phase—extracted from two-site reduced density matrices of the transverse-field Ising model with nearest-neighbor interacting spins. By applying coordinated local unitary rotations to the spins, we compute the geometric phases associated with the two-site states across the critical point. We find that the interferometric phase is a more reliable indicator of the quantum phase transition in this model than the Uhlmann phase.
\end{abstract}
\maketitle

\section{Introduction}

Geometric phases (GPs) of mixed quantum states \cite{uhlmann86,sjoqvist00} have attracted significant interest due to their fundamental and practical implications in quantum physics \cite{bhandari02,anandan02,filipp03,du03,levay04,sjoqvist04,filipp05,ericsson05,sarandy06,milman06,klepp08,martin13,viyuela18,he22,lombardo26,yang26}. 
For adiabatic evolution in the pure state limit, GP is uniquely known as the Berry phase \cite{berry84}, whereas for mixed states distinct underlying geometric structures are reflected by the different forms of GP \cite{slater02,ericsson03,rezakhani06,zhu11,andersson15,kiselev18}. The behavior near quantum critical points has been studied for pure states \cite{carollo05,zhu06,plastina06,azimi13,carollo20} and for many-body systems at finite temperatures \cite{viyuela14,huang14,budich15,andersson16}. While these previous studies on quantum criticality have focused on the GP of the global state, less attention has been paid to GPs associated with reduced density matrices, which describe local subsystems in a correlated many-body environment, possibly at zero temperature.

Reduced states are of particular interest in strongly interacting systems, because even when the global state is pure, subsystem states are typically mixed due to entanglement. The GP of a reduced density matrix therefore contains information not only about local structures, but also about non-local correlations inherited from the full system \cite{osborne02,dillenschneider08,sultan25}. 

Here, we consider the transverse-field Ising model, one of the most extensively studied exactly solvable models of quantum critical behavior \cite{sachdev99}. The model exhibits a second-order quantum phase transition (QPT) at the critical point $\lambda_c = 1$, separating the ferromagnetic and paramagnetic phases. Since it is exactly solvable, it allows for analytic expressions of correlation functions and reduced density matrices in the thermodynamic limit.

Rather than studying the GP of the full many-body ground state, we focus on two-site reduced density matrices extracted from the chain. These states are generally mixed due to entanglement with the rest of the system and depend explicitly on spin–spin correlation functions. By applying global adiabatic single-spin rotations forming loops in parameter space, we evaluate the interferometric \cite{sjoqvist00} and Uhlmann  \cite{uhlmann86} GPs of the two-spin reduced state as well as the deviation of these phases from the GP of single-spin reduced state. We in particular investigate the dependence on inter-site distance of these two forms of mixed state GPs at zero temperature $T$. The purpose of the analysis is to determine whether the two notions of GP can be used as order parameter that distinguishes the phases on either side of the quantum crtitical point of the Ising chain. 

The paper is organized as follows. In Sec.~\ref{sec:model}, we briefly review the transverse-field Ising model, identify the reduced two-spin density matrices, and describe the adiabatic path generating the GPs. In Sec.~\ref{sec:interferometric}, we compute the interferometric GP and analyze its dependence on the transverse field and inter-site distance. We evaluate the Uhlmann GP and compare its behavior with the interferometric one in Secs.~\ref{sec:uhlmann} and \ref{sec:compare}. The paper ends with the conclusions. 

\section{Setup}
\label{sec:model}

For a chain of $N$ nearest-neighbor interacting spins, the transverse-field Ising model is given by 
\begin{eqnarray}
H = - \lambda \sum_{j=1}^N \sigma_j^x \sigma_{j+1}^x - \sum_{j=1}^N \sigma_j^z ,
\end{eqnarray}
where $\lambda$ parametrizes the ratio between the interaction and transverse-field strengths. We assume  periodic boundary conditions, such that $\sigma_{N+1}^{\alpha} \equiv \sigma_1^{\alpha}$, $\alpha=x,y,z$. The system exhibits a QPT at $\lambda_c = 1$, separating a ferromagnetic ($\lambda > 1$) from a paramagnetic ($\lambda < 1$) phase. The reduced two-spin density matrices contain well-defined mixed states whose GPs will be investigated in the following sections. 

The model can be diagonalized exactly by first mapping the spin operators to spinless fermionic operators through the Jordan–Wigner transformation. This transformation converts the spin Hamiltonian into a quadratic fermionic Hamiltonian. Subsequently, a Fourier transformation followed by a Bogoliubov transformation are applied, which diagonalize the Hamiltonian in terms of noninteracting fermionic quasiparticles $\eta_q$. This results in 
\begin{eqnarray}
H = 2 \sum_q \omega_q \eta_q^\dagger \eta_q 
- \sum_q \omega_q,
\end{eqnarray}
with dispersion relation
\begin{eqnarray}
\omega_{\phi_q} = \sqrt{(\lambda \sin \phi_q)^2 + (1 + \lambda \cos \phi_q)^2 },
\qquad \phi_q = \frac{2\pi q}{N}.
\end{eqnarray}

At thermal equilibrium, the spin chain is described by the Gibbs state $\rho = e^{-\beta H}/Z$, where $\beta$ is the inverse temperature and $Z={\rm Tr} \left( e^{-\beta H} \right)$ is the partition function. The symmetries of the system allow us to express the one- and two-spin reduced density matrices entirely in terms of the magnetization $\langle \sigma^z \rangle$ and two-site correlation functions $\langle \sigma_0^{\alpha} \sigma_r^{\alpha} \rangle$ for spins at sites $j,k$, such that $|j-k|=r$. Here, we focus on the low temperature limit $T \xrightarrow{} 0$, in which the parameter $\beta \rightarrow \infty$ and only the ground state will be populated. In the thermodynamic limit $N \rightarrow \infty$ at $T = 0\,\mathrm{K}$, the magnetization is given by
\begin{eqnarray}
\langle \sigma^z \rangle = \frac{1}{\pi} \int_0^\pi \frac{1 + \lambda \cos\phi}{\omega_\phi} d\phi,
\label{eq:magnetization}
\end{eqnarray} 
where we have replaced $\phi_q$ by the continuous variable $\phi$. The two–site correlation functions can be expressed in terms of the Toeplitz matrix elements  
\begin{eqnarray}
G_r & = &
\frac{1}{\pi} 
\int_0^\pi d\phi\;
\cos(r\phi)\,
\frac{1 + \lambda \cos\phi}{\omega_\phi} 
\nonumber \\
& & + \frac{\lambda}{\pi} \int_0^\pi d\phi 
\sin(r\phi)\, \frac{\sin\phi}{\omega_\phi}.
\label{eq:Gr}
\end{eqnarray}
The transverse correlations in the $x$ direction are given by
\begin{eqnarray}
\langle \sigma_0^x \sigma_r^x \rangle =
\begin{vmatrix}
G_{-1} & G_{-2} & \cdots & G_{-r} \\
G_{0}  & G_{-1} & \cdots & G_{-r+1} \\
\vdots & \vdots & \ddots & \vdots \\
G_{r-2} & G_{r-3} & \cdots & G_{-1}
\end{vmatrix} .
\label{eq:xcorrelation}
\end{eqnarray}
Similarly, the transverse $y$ correlations are
\begin{eqnarray}
\langle \sigma_0^y \sigma_r^y \rangle =
\begin{vmatrix}
G_{1} & G_{0}  & \cdots & G_{-r+2} \\
G_{2} & G_{1}  & \cdots & G_{-r+3} \\
\vdots & \vdots & \ddots & \vdots \\
G_{r} & G_{r-1} & \cdots & G_{1}
\end{vmatrix},
\label{eq:ycorrelation}
\end{eqnarray}
which has the same Toeplitz determinant structure but with indices
shifted according to the fermionic contractions.
The longitudinal $z$ correlations are given by
\begin{eqnarray}
\langle \sigma_0^z \sigma_r^z \rangle
= \langle \sigma^z \rangle^2 - G_r G_{-r} .
\label{eq:zcorrelation}
\end{eqnarray}
The reduced one- and two-spin density matrices read \cite{osborne02}  
\begin{eqnarray}
\rho_1 = \frac{1}{2}\big( \mathbb{I} + \langle \sigma^z \rangle \sigma^z \big), 
\label{eq:sing_site}
\end{eqnarray}
and 
\begin{eqnarray}
\rho_{0r} = \frac14 \left(
\mathbb{I} \otimes \mathbb{I}
+ \langle \sigma^z\rangle(\sigma_0^z + \sigma_r^z)
+ \sum_{\alpha=x,y,z} \langle \sigma_0^{\alpha} \sigma_r^{\alpha} \rangle \,
\sigma_0^{\alpha} \sigma_r^{\alpha}
\right) , 
\end{eqnarray}
respectively. 

We consider adiabatic paths that simultaneously rotate the local spins around $y$ and $z$ axes by angles $\theta$ and $\varphi$, respectively. This corresponds to the unitary
\begin{eqnarray}
 U_1(\varphi,\theta) & = & R_z(\varphi) R_y(\theta) 
 \nonumber \\ 
  & = &  
\begin{pmatrix}
e^{i\varphi/2} & 0\\
0 & e^{-i\varphi/2}
\end{pmatrix} 
\begin{pmatrix}
\cos \frac{\theta}{2} & -\sin \frac{\theta}{2}\\
\sin \frac{\theta}{2} & \cos \frac{\theta}{2}
\end{pmatrix} ,   
\end{eqnarray}
yielding the paths 
\begin{eqnarray}
\rho_1 \mapsto \rho_1 (\varphi,\theta) & = &   U_1(\varphi,\theta) \rho_1  U_1^{\dagger}(\varphi,\theta) , 
\nonumber \\ 
\rho_{0r} \mapsto \rho_{0r}(\varphi,\theta) & = &   U_2(\varphi,\theta) \rho_{0r} U_2^{\dagger}(\varphi,\theta) , 
\end{eqnarray}
with $U_2(\varphi,\theta)$ short-hand notation for $U_1(\varphi,\theta) \otimes U_1(\varphi,\theta)$.  Being locally applied to each spin, all correlations in the system will remain unchanged under the rotation. 

\section{Interferometric GP}
\label{sec:interferometric}

\begin{figure*}[t]
\centering
\includegraphics[width=0.49\textwidth,height=0.37\textheight]{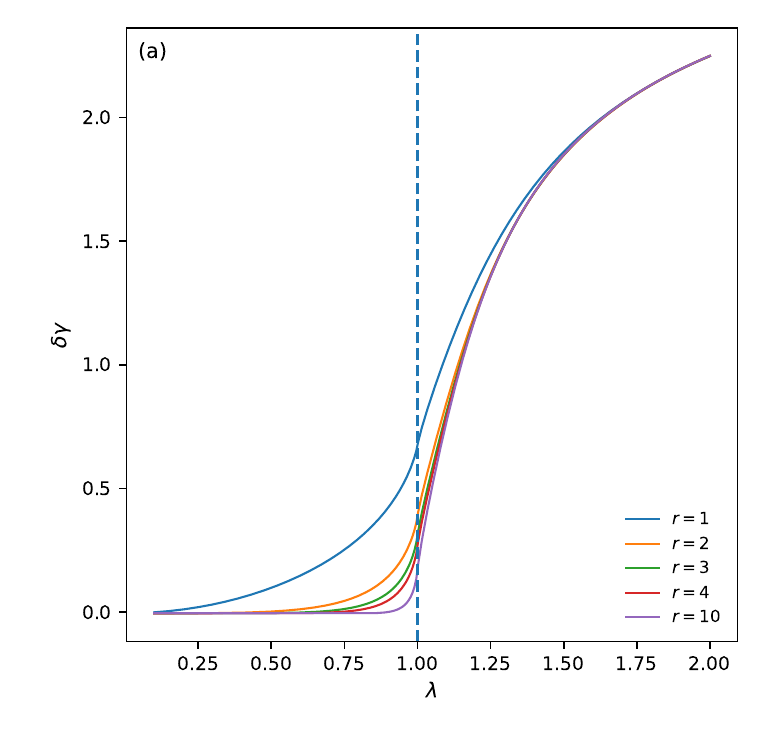}
\hfill
\includegraphics[width=0.49\textwidth,height=0.37\textheight]{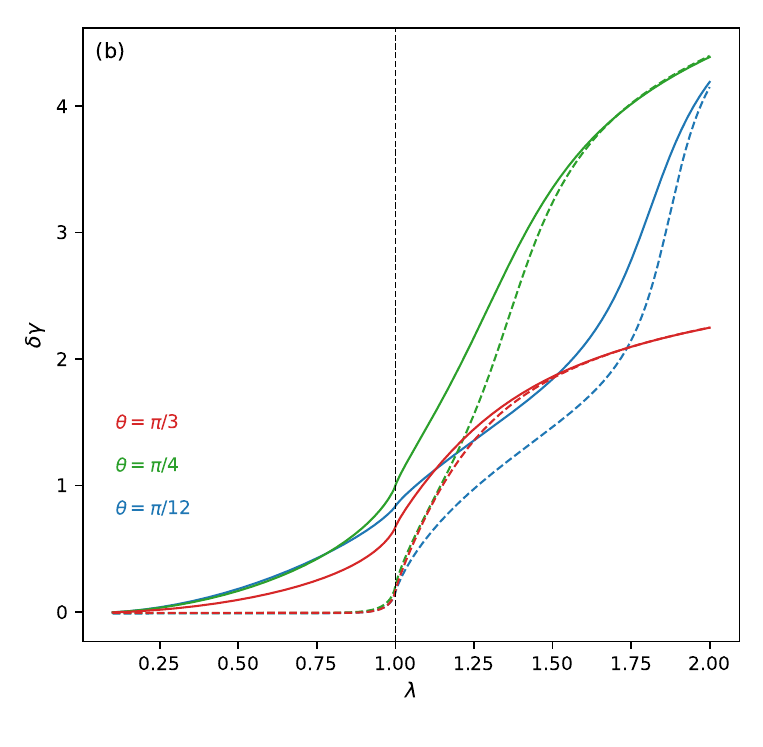}
\caption{(a) shows the interferometric GP deviation $\delta\gamma(r, \theta) \equiv \delta \gamma$ measured on the two-site reduced density matrix of the transverse-field Ising chain as a function of the control parameter $\lambda$ at $\theta = \frac{\pi}{3}$ for various $r$. (b) shows $\delta \gamma$ as a function of $\lambda$ for $\theta = \frac{\pi}{3},\frac{\pi}{4},\frac{\pi}{12}$. The solid lines correspond to nearest-neighbor sites ($r=1$) and the dashed lines represent $r=10$.}
\label{fig:results_1}
\end{figure*}

Let the initial two-spin density matrix have the spectral decomposition
\begin{eqnarray}
\rho_{0r} = \sum_n p_n \ket{n} \bra{n},     
\label{eq:rho_0r}
\end{eqnarray}
with $p_n\ge 0$, $\sum_n p_n =1$, and $\langle n \ket{n'} = \delta_{nn'}$. By rotating the system, these eigenstates evolve as $\ket{n} \mapsto \ket{n(\varphi,\theta)} = U_2(\varphi,\theta) \ket{n}$ with all eigenvalues $p_n$ being constant. We consider adiabatic loops at fixed $\theta$, resulting in the interferometric GP \cite{sjoqvist00}  
\begin{eqnarray}
\gamma (\rho_{0r},\theta) & = & \arg \left[
\sum_{n} p_{n} \langle n(0; \theta)\ket{n(2\pi; \theta)} \right.  
\nonumber \\
 & & \left. \times \exp \left( -\int_{0}^{2\pi} \langle n(\varphi; \theta) \ket{\partial_{\varphi}n (\varphi; \theta)} d\varphi \right) \right] ,  
\label{eq:gamma_int}
\end{eqnarray}
where we have used the notation $(\varphi;\theta)$ to indicate that $\theta$ is a fixed parameter controlling the path dependence. 

To analyze the behavior of the two-site interferometric GP, we compare the phase of the two-spin density matrix $\rho_{0r}$ with that of the corresponding single-spin reduced density matrix $\rho_1$. In the limit of large inter-site separation $r$, spin-spin correlations may decay, in case of which the two-spin reduced state approaches a product form: 
\begin{eqnarray}
\rho_{0r} \longrightarrow \rho_1 \otimes \rho_1 .
\end{eqnarray}
In this asymptotic regime, the GP of the two-spin state is
\begin{eqnarray}
\gamma(\rho_{0r}) \longrightarrow \gamma(\rho_1 \otimes \rho_1) = 2\,\gamma(\rho_1) . 
\end{eqnarray}
Motivated by this limiting behavior, we introduce the quantity \cite{williamson07}
\begin{eqnarray}
\delta\gamma(r,\theta) =  \gamma(\rho_{0r},\theta) - 2\gamma (\rho_1,\theta) , 
\label{eq: 19}
\end{eqnarray}
which quantifies the deviation of the two-spin GP from its product state value. Note that, the GP of the single-spin reduced density matrix in Eq.~\eqref{eq:sing_site} becomes \cite{sjoqvist00} 
\begin{eqnarray}
\gamma (\rho_1,\theta) = -\arctan \left( \langle \sigma_z \rangle 
\tan \frac{\Omega}{2} \right)  
\label{eq:gamma_int}
\end{eqnarray}
with $\Omega = 2\pi (1-\cos\theta)$ the enclosed solid angle on the parameter sphere and $\langle \sigma_z \rangle$ given by Eq.~\eqref{eq:magnetization}. 

Figure \ref{fig:results_1} shows the interferometric GP deviation $\delta\gamma(r, \theta) \equiv \delta\gamma$ as a function of the control parameter $\lambda$. The results indicate that $\delta\gamma$ behaves as an effective order parameter capable of distinguishing the different phases of the transverse-field Ising chain and signaling the onset of the QPT at $\lambda_c=1$. Near the critical point, $\delta\gamma$ undergoes a rapid yet continuous change, consistent with the critical behavior of a continuous (second-order) phase transition. The progressively steeper variation of the curves across $\lambda_c$ suggests the development of a strongly enhanced, potentially divergent response. For $\lambda<1$, $\delta\gamma$ remains strongly suppressed for all angles $\theta$ [see Fig.~\ref{fig:results_1} (b)], whereas above criticality it increases continuously and the curves tend to converge for different values of $r$. Increasing $r$ leads to a stronger suppression of $\delta\gamma$ in the paramagnetic phase thereby sharpening the crossover at  $\lambda_c$. This behavior seems to reflect the fact that, as the distance between spins increases, their mutual correlations become negligible below the critical point, while they persist in the ferromagnetic phase \cite{pfeuty70}. 

\section{Uhlmann Phase}
\label{sec:uhlmann}

\begin{figure*}[t]
\centering
\includegraphics[width=0.49\textwidth, height=0.37\textheight]{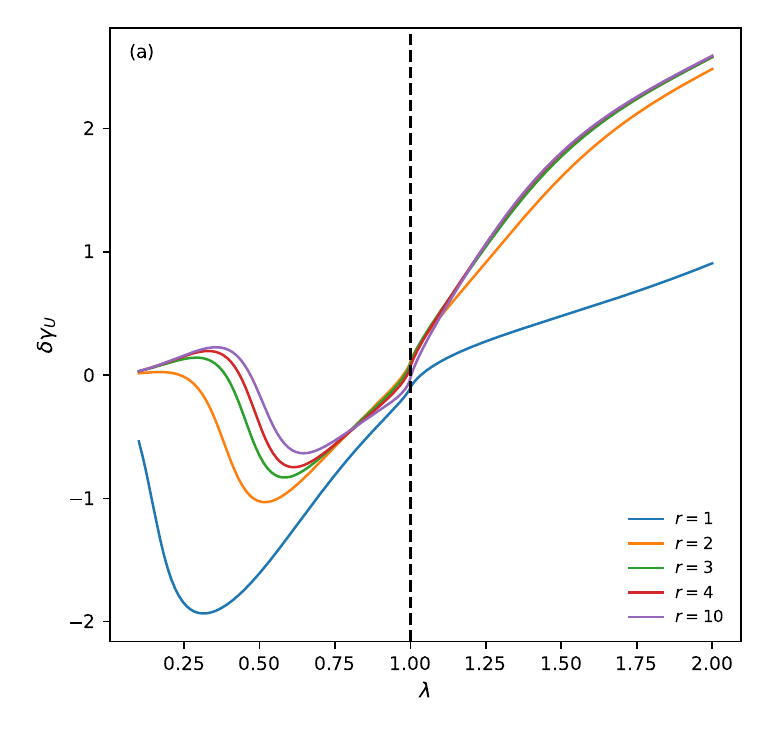}
\hfill
\includegraphics[width=0.49\textwidth,height=0.37\textheight]{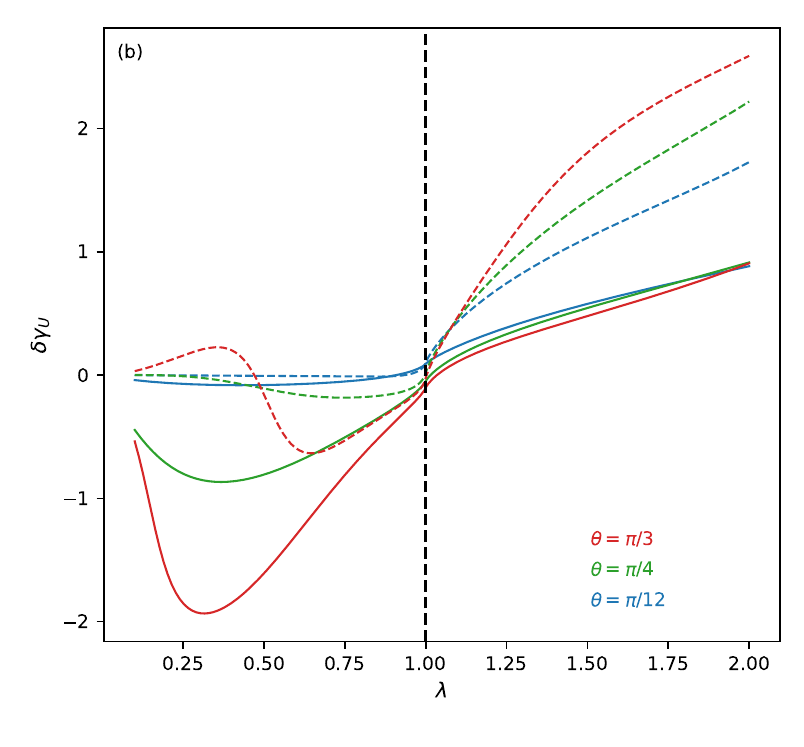}
\caption{(a) shows the Uhlmann phase deviation $\delta \gamma_U (r, \theta) \equiv \delta \gamma_U$ measured on the two-site reduced density matrix of the transverse-field Ising chain as a function of the control parameter $\lambda$ at $\theta = \frac{\pi}{3}$ for various $r$. (b) shows $\delta \gamma_U$ as a function of $\lambda$ for $\theta = \frac{\pi}{3},\frac{\pi}{4},\frac{\pi}{12}$. The solid line is corresponding to the $r=1$ and the dashed line is representing $r=10$.}
\label{fig:results_4}
\end{figure*}

The Uhlmann phase is defined using purification method, in which the density matrix $\rho$ acting on Hilbert space $\mathcal{H}$ is represented by a pure state in some extended Hilbert-space $\mathcal{H} \otimes \mathcal{H}_a$. Following Uhlmann, a purification (or `amplitude') of $\rho$ is a Hilbert–Schmidt operator $w: \mathcal{H}_a \mapsto \mathcal{H}$ satisfying
\begin{eqnarray}
\rho = w w^\dagger, 
\end{eqnarray}
where $\mathcal{H}_a$ is an auxiliary Hilbert space. The purification takes the form 
\begin{eqnarray}
w = \sqrt{\rho} \, V 
\end{eqnarray}
with $V$ an arbitrary unitary (or `phase factor') acting on the auxiliary system $a$ \cite{ericsson03}. Purifications $w$ and $\tilde{w}$ of density matrices $\rho$ and $\tilde{\rho}$, respectively, both assumed to be full rank, are said to be parallel or `in-phase' if 
\begin{eqnarray}
\tilde{w}^{\dagger} w > 0 .  
\end{eqnarray} 

We again restrict to loops where $\theta$ remains constant and $\varphi$ varies from $0$ to $2\pi$. In this case, the Uhlmann parallel transport condition reads 
\begin{eqnarray}
w^\dagger (\varphi;\theta) \partial_{\varphi} w(\varphi;\theta) = \partial_{\varphi} w^{\dagger} (\varphi;\theta) w(\varphi;\theta) .
\end{eqnarray}
This implies that $V(\varphi;\theta)$ evolves according to
\begin{eqnarray}
\partial_{\varphi} V(\varphi;\theta) = A(\varphi;\theta)  V(\varphi;\theta)  
\end{eqnarray}
where $A(\varphi;\theta)$ is the anti-Hermitian Uhlmann connection with matrix elements \cite{hubner93} 
\begin{eqnarray}
A_{nm} (\varphi;\theta) = 
\frac{\bra{n (\varphi;\theta)} \left[ \partial_{\varphi} \sqrt{\rho_{0r} (\varphi;\theta)},\sqrt{\rho_{0r} (\varphi;\theta)} \right] 
\ket{m (\varphi;\theta)}}{p_n + p_m}  
\nonumber \\ 
\label{eq:Huebner_connection}
\end{eqnarray}
with $\rho_{0r} (\varphi;\theta) = U_2 (\varphi;\theta) \rho_{0r} U_2^{\dagger} (\varphi;\theta)$. The unitary $V(\varphi;\theta)$ is propagated along the path using
\begin{eqnarray}
V_{k+1} = \exp \big[ A(\varphi_k;\theta) \delta\varphi \big] \, V_k, \qquad V_0 = I,
\end{eqnarray} 
with $\varphi_{k+1} = \varphi_k + \delta \varphi$. 
The Uhlmann phase is obtained as the relative phase of the initial [$w(0;\theta)$] and final [$w(2\pi;\theta)$] amplitudes:
\begin{eqnarray}
\gamma_U (\rho_{0r},\theta) & = & \arg \operatorname{Tr}
\big[ w^{\dagger} (0;\theta) w(2\pi;\theta) \big] 
\nonumber \\ 
 & = & \arg \operatorname{Tr} \big[ \rho_{0r}(0;\theta) V(2\pi;\theta) \big] , 
\label{eq:Uhlmann_phase}
\end{eqnarray}
where we have used that $w(0;\theta) = \sqrt{\rho(0;\theta)}$ and $w(2\pi;\theta) = \sqrt{\rho(0;\theta)} V(2\pi;\theta)$ for a loop. 
This expression captures the full mixed-state GP around the loop. Similar to the interferometric case, we introduce the quantity \cite{williamson07}
\begin{eqnarray}
\delta\gamma_U (r,\theta) =  \gamma_U (\rho_{0r},\theta) - 2\gamma_U (\rho_1,\theta)  
\label{eq: 19}
\end{eqnarray}
with $\gamma_U(\rho_1,\theta)$ the single-site Uhlmann phase. This quantifies the deviation of the two-spin Uhlmann phase from its product state value.

Figure \ref{fig:results_4} displays the Uhlmann phase deviation $\delta\gamma_U (r,\theta) \equiv \delta\gamma_U$ as a function of the control parameter $\lambda$. The results show that the behavior of $\delta\gamma_U$ strongly depends on both the spin separation $r$ and the path parameter $\theta$ even in the paramagnetic phase. Although $\delta\gamma_U$ does not universally exhibit a sharp singularity or extremum at the critical point $\lambda_c = 1$, the curves display a noticeable compression, indicating that the mixed-state geometric structure picked up by $\delta \gamma_U$ is influenced by the critical reorganization of the many-body ground state. As $\lambda \to 0$, the Uhlmann phase deviation tends to zero, consistent with the product state limit in the paramagnetic phase. This behavior is clearly visible in the figure, although the convergence is considerably slower and more sensitive to $\theta$ than in the interferometric GP case. 

\section{Behavior near the critical point}
\label{sec:compare}

\begin{figure*}[t]
\centering
\includegraphics[width=0.49\textwidth, height=0.37\textheight]{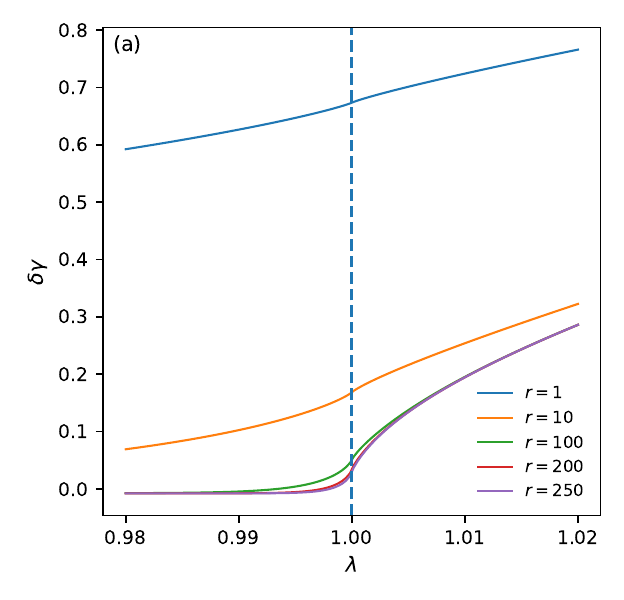}
\hfill
\includegraphics[width=0.49\textwidth,height=0.368\textheight]{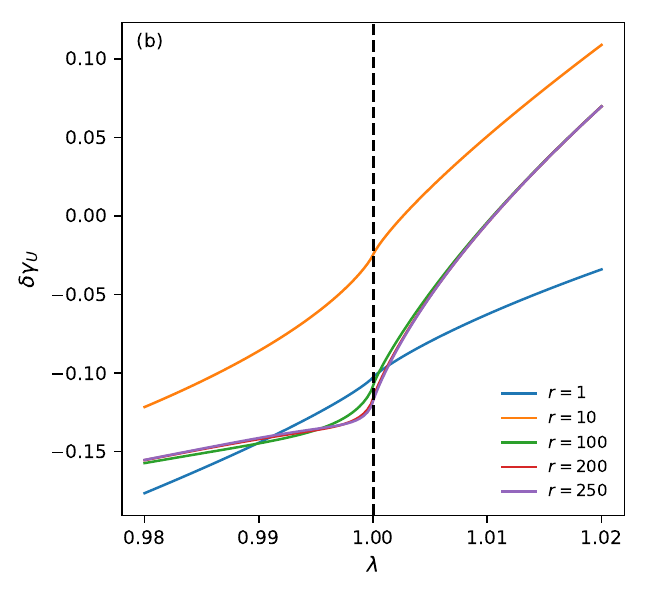}
\caption{(a) shows the interferometric GP deviation $\delta\gamma(r, \theta) \equiv \delta \gamma$ and (b) shows the Uhlmann phase deviation $\delta \gamma_U (r, \theta) \equiv \delta \gamma_U$, measured on the two-site reduced density matrix of the transverse-field Ising chain. The two GPs are shown as functions of the control parameter $\lambda$ at $\theta = \frac{\pi}{3}$ for various $r$ near critical point $\lambda_c = 1$.}
\label{fig:results_5}
\end{figure*}

Figure \ref{fig:results_5} presents a comparison between the interferometric GP deviation $\delta\gamma$ and the Uhlmann phase deviation $\delta\gamma_U$ near the quantum critical point $\lambda_c = 1$ for different inter-site distances $r$. It is evident that both GPs capture the underlying quantum critical properties of the system. The interferometric GP deviation shown in  Fig. \ref{fig:results_5} (a) varies smoothly across the transition and exhibits a gradual enhancement as the control parameter approaches the critical region, with the response becoming increasingly pronounced for larger values of $r$. In contrast, the Uhlmann phase deviation in Fig. \ref{fig:results_5} (b) shows a larger slope across $\lambda_c$. More importantly, $\delta \gamma$ 
tends to vanish below the critical point as $r$ increases, clearly reflecting the absence of long-range correlations in the paramagnetic phase. On the other hand, $\delta\gamma_U$ does not vanish for $\lambda \leq 1$ even for very large $r$. One possible explanation for this behavior is the presence of the auxiliary physical system required to implement the Uhlmann parallel transport \cite{ericsson03}. This construction may introduce additional correlations, which could account for the nonvanishing behavior of $\delta\gamma_U$ in the limit of large inter-site distances within the paramagnetic phase.

\section{Conclusions}

We have carried out a comparative study of the interferometric GP and the Uhlmann phase for two-spin reduced states in the transverse-field Ising chain. Their quantitative behavior differs, highlighting the inequivalent geometric structures underlying the two forms of mixed-state GP. The interferometric GP reflects an operationally defined geometric interference effect, whereas the Uhlmann phase encodes parallel transport in the space of purifications. Their comparison therefore sheds light on how different notions of mixed-state geometry respond to many-body critical fluctuations. In particular, compared with the interferometric GP, the critical signatures extracted from the Uhlmann phase appear less universal and do not capture the absence of long-range correlations in the paramagnetic phase. Additional support for the generality of these findings would require extending the analysis to systems with longer-range interactions and to other quantum spin models.

\section*{Acknowledgements}
E. S. acknowledges financial support from the Swedish Research Council (VR) through Grant No. 2025-05249.

\end{document}